\documentclass[11pt,a4paper]{article}
\usepackage{jheppub_kim}

\usepackage{pdflscape}
\usepackage{amsmath}
\usepackage{amssymb}
\usepackage{dcolumn}
\usepackage{bm}
\usepackage{color}
\usepackage{epsfig}
\usepackage{amsfonts}
\usepackage{graphicx}
\usepackage{subfigure}
\usepackage{dcolumn}

\newcommand{\be}{\begin{equation}}
\newcommand{\ee}{\end{equation}}
\newcommand{\bea}{\begin{eqnarray}}
\newcommand{\eea}{\end{eqnarray}}
\newcommand{\nn}{\nonumber}

\renewcommand{\[}{\left[}

\def\be{\begin{equation}}
\def\ee{\end{equation}}
\def\bea{\begin{eqnarray}}
\def\eea{\end{eqnarray}}

\begin{document}

\title{Dynamical behavior in $f(T,T_G)$ cosmology}

\author[a]{Georgios Kofinas}

\author[b]{Genly Leon}

\author[c,b]{Emmanuel N. Saridakis}

\affiliation[a]{Research Group of Geometry, Dynamical Systems and
Cosmology, Department of Information and Communication Systems Engineering,
University of the Aegean, Karlovassi 83200, Samos, Greece}

\affiliation[b]{Instituto de F\'{\i}sica, Pontificia Universidad  Cat\'olica
de Valpara\'{\i}so, Casilla 4950, Valpara\'{\i}so, Chile}

\affiliation[c]{Physics Division, National Technical University of Athens,
15780 Zografou Campus,  Athens, Greece}

\emailAdd{gkofinas@aegean.gr}

\emailAdd{genly.leon@ucv.cl}

\emailAdd{Emmanuel$_-$Saridakis@baylor.edu}


\abstract{The $f(T,T_G)$ class of gravitational modification, based on the
quadratic torsion scalar $T$, as well as on the new quartic torsion scalar
$T_G$ which is the teleparallel equivalent of the Gauss-Bonnet term, is a
novel theory, different from both $f(T)$ and $f(R,G)$ ones. We perform a
detailed dynamical analysis of a spatially flat universe governed by the
simplest non-trivial model of $f(T,T_G)$ gravity which does not introduce a
new mass scale.
We find that the universe can result in dark-energy dominated, quintessence-like,
cosmological-constant-like or phantom-like solutions, according to the
parameter choices. Additionally, it may result to a  dark energy - dark
matter scaling solution, and thus it can alleviate the coincidence problem.
Finally, the analysis ``at infinity'' reveals that the universe may exhibit
future, past, or intermediate singularities depending on the parameters.}

\keywords{Modified gravity, dark energy, Gauss-Bonnet, $f(T)$ gravity,
 dynamical analysis}

\maketitle


\section{Introduction}
\label{Introduction}

Since the discovery of the universe late-times acceleration, a large amount
of research has been devoted to its explanation. In principle, one can follow
two main directions to achieve this. The first way is to modify the content
of the universe introducing the dark energy concept, with its simpler
candidates being a canonical scalar field, a phantom field or the
combination of both fields in a unified model dubbed quintom (for reviews on
dark energy see \cite{Copeland:2006wr,Cai:2009zp} and references
therein). The second direction that one can follow is to modify the
gravitational sector itself (for a review see \cite{Capozziello:2011et} and
references therein), acquiring a modified cosmological dynamics. However,
note that apart from the interpretation, one can transform from one approach
to the other, since the crucial issue is just the number of degrees of
freedom beyond General Relativity and standard model particles (see
\cite{Sahni:2006pa} for a review on such a unified point of view). Finally,
note that the above scenarios, apart from late-times implications, can be
also used for the description of the inflationary stage \cite{Nojiri:2003ft}.

In the majority of modified gravitational theories, one suitably extends the
curvature-based Einstein-Hilbert action of General Relativity. However, an
interesting class of gravitational modification arises when one modifies the
action of the equivalent formulation of General Relativity based on torsion.
In particular, it is known that Einstein himself constructed the so-called
``Teleparallel Equivalent of General Relativity'' (TEGR)
\cite{Unzicker:2005in,Hayashi:1979qx,Pereira,Maluf:2013gaa} using the
curvature-less
Weitzenb{\"{o}}ck connection instead of the torsion-less Levi-Civita one.
The corresponding Lagrangian, namely the torsion scalar $T$, is constructed
by contractions of the torsion tensor, in a similar way that the usual
Einstein-Hilbert Lagrangian $R$ is constructed by contractions of the
curvature (Riemann) tensor. Thus, inspired by the $f(R)$ modifications of
the Einstein-Hilbert Lagrangian \cite{DeFelice:2010aj,Nojiri:2010wj}, one can
construct the $f(T)$ modified gravity by extending $T$ to an arbitrary
function \cite{Ferraro:2006jd,Ben09,Linder:2010py}. Note that although TEGR
coincides with General Relativity at the level of equations, $f(T)$ does not
coincide with $f(R)$, that is they represent different modification classes.
Thus, the cosmological implications of $f(T)$ gravity are new and
interesting
\cite{Linder:2010py,Chen:2010va,Dent:2011zz,Zheng:2010am,
Sharif001,Li:2011rn, Cai:2011tc,Boehmer:2011gw,Capozziello:2011hj,
Daouda:2011rt,Geng:2011aj,Wu:2011kh,Gonzalez:2011dr,Wei:2011aa,
Atazadeh:2011aa,
Farajollahi:2011af,Karami:2012fu,Iorio:2012cm,Cardone:2012xq,
Capozziello:2012zj,Jamil:2012ti,
Ong:2013qja, Amoros:2013nxa, Otalora:2013dsa, Geng:2013uga,Nesseris:2013jea,
Bamba:2013ooa,
Nashed:2014uta,Harko:2014sja,Harko:2014aja}.

However, in curvature gravity, apart from the simple $f(R)$ modification one
can construct more complicated extensions using higher-curvature corrections
such as the Gauss-Bonnet term $G$
\cite{Wheeler:1985nh,Antoniadis:1993jc,Nojiri:2005vv} or
functions of it
  \cite{Nojiri:2005jg,DeFelice:2008wz}, Lovelock combinations
\cite{Lovelock:1971yv,Deruelle:1989fj,Charmousis:2008kc}, and Weyl
combinations
\cite{Mannheim:1988dj,Flanagan:2006ra,Grumiller:2013mxa}. Inspired
by these, in the recent work \cite{Kofinas:2014owa}, the $f(T,T_G)$
gravitational modification was constructed, which is based on the
old quadratic torsion scalar $T$, as well as on the new quartic
torsion scalar $T_G$ that is the teleparallel equivalent of the
Gauss-Bonnet term. Obviously, $f(T,T_G)$ theories cannot arise from
the $f(T)$ ones, and additionally they are different from $f(R,G)$
class of curvature modified gravity. Thus,  $f(T,T_G)$ is a novel
class of gravitational modification.

The cosmological applications of $f(T,T_G)$ gravity proves to be very
interesting \cite{Kofinas:2014owa}. Therefore, it is both interesting and
necessary to perform a dynamical analysis, examining in a systematic way the
allowed cosmological behaviors, focusing on the late-times stable solutions.
The phase-space and stability analysis is a very powerful tool, since it
reveals the global features of a given cosmological scenario,
independently of the initial conditions and the specific evolution of the
universe. In the present investigation we perform such a detailed
phase-space analysis, and we extract the late-times, asymptotic solutions,
calculating also the corresponding observable quantities, such as the
deceleration parameter, the effective dark energy equation-of-state
parameter, and the various density parameters.

The plan of the work is the following: In section \ref{fTgravity}
we briefly review the scenario of $f(T,T_G)$ gravity and in section
\ref{fTcosmology} we present its application in cosmology. In section
\ref{Phasespaceanalysis} we perform the detailed dynamical analysis for the simplest
non-trivial model of $f(T,T_{G})$ gravity.
In section \ref{implications} we discuss the cosmological implications and the
physical behavior of the scenario. Finally, in section \ref{Conclusions} we
summarize our results.

\section{$f(T,T_G)$ gravity}
\label{fTgravity}

In this section we briefly review the $f(T,T_G)$  gravitational modification
following \cite{Kofinas:2014owa}.
In the whole manuscript we use the following
notation: Greek indices   run over the coordinate
space-time, while Latin indices run over its tangent space.

In this framework the dynamical variable is the vierbein field $e_a(x^\mu)$.
In terms of coordinates, it can be expressed in components as
$e_a=e^{\,\,\, \mu}_a\partial_\mu$, while the dual vierbein is defined as
$e^a=e^a_{\,\,\, \mu}d x^\mu$. Concerning the other field, that is the
connection   1-forms $\omega^a_{\,\,\,
b}(x^\mu)$ which defines the parallel transportation, one uses the
Weitzenb{\"{o}}ck one, which in all coordinate frames is defined as
\begin{eqnarray}
\omega_{\,\,\,\mu\nu}^{\lambda}=e_{a}^{\,\,\,\lambda}e^{a}_{\,\,\,\mu
, \nu } .
\label{Weinzdef}
\end{eqnarray}
Due to its inhomogeneous transformation law it has tangent-space
components $\omega_{\,\,\,bc}^{a}=0$, assuring the property of vanishing non-metricity.
Additionally, for an orthonormal vierbein the metric tensor is given by the relation
\begin{equation}
\label{metrdef}
g_{\mu\nu} =\eta_{ab}\, e^a_{\,\,\,\mu}  \, e^b_{\,\,\,\nu},
\end{equation}
where $\eta_{ab}=\text{diag}(-1,1,1,1)$ and indices $a,b,...$ are
raised/lowered with $\eta_{ab}$.

One can now define the torsion  tensor as
\begin{equation}
T^{\lambda}_{\,\,\,\mu\nu}=
e_{a}^{\,\,\,\lambda}\left(\partial_\nu e^{a}_{\,\,\,\mu}-\partial_\mu
e^{a}_{\,\,\,\nu}\right),
\end{equation}
while the Riemann tensor is zero by construction, due to the
teleparallelism condition which is imposed with the use of the
Weitzenb{\"{o}}ck  connection.
Moreover, the contorsion tensor, which equals the difference
between the Weitzenb%
\"{o}ck and Levi-Civita connections, is defined as
\begin{equation}  \label{cotorsion}
\mathcal{K}^{\mu\nu}_{\:\:\:\:\rho}=-\frac{1}{2}\Big(T^{\mu\nu}_{
\:\:\:\:\rho}
-T^{\nu\mu}_{\:\:\:\:\:\rho}-T_{\rho}^{\:\:\mu\nu}\Big).
\end{equation}

Since in this formulation all the information concerning the
gravitational field is included in the torsion tensor
$T^{\lambda}_{\,\,\,\mu\nu}$, one can use it in order to construct torsion
invariants.
The simplest invariants that one can build are quadratic in the torsion
tensor. In particular, the combination
\begin{eqnarray}
T&=&\frac{1}{4}T^{\mu\nu\lambda}T_{\mu\nu\lambda}+\frac{1}{2}T^{\mu\nu\lambda}
T_{\lambda\nu\mu}-T_{\nu}^{\,\,\,\nu\mu}T^{\lambda}_{\,\,\,\lambda\mu},
\label{Tquad}
\end{eqnarray}
which can in general be defined in an arbitrary dimension $D$, is the ``torsion scalar'',
and if it is used as a Lagrangian and be varied
in terms of the vierbein it gives rise to the Einstein field equations.
That is why the gravitational theory characterized by the action
\begin{eqnarray}
S=-\frac{1}{2\kappa^2}\int d^4 x \,e\,T \, ,
\label{teleaction}
\end{eqnarray}
with $e=\det{(e^{a}_{\,\,\,\mu})}=\sqrt{|g|}$ and $\kappa^2\equiv 8\pi
G$  the   gravitational constant, is called Teleparallel
Equivalent of General Relativity. In these lines, one can be based
on $T$ in order to construct modified gravitational theories extending the
TEGR action to
\cite{Linder:2010py,Chen:2010va,Dent:2011zz,Zheng:2010am,
Sharif001,Li:2011rn,
Cai:2011tc,Boehmer:2011gw,Capozziello:2011hj,
Daouda:2011rt,Geng:2011aj,Wu:2011kh,Gonzalez:2011dr,Wei:2011aa,
Atazadeh:2011aa,
Farajollahi:2011af,Karami:2012fu,Iorio:2012cm,Cardone:2012xq,
Capozziello:2012zj,Jamil:2012ti,
Ong:2013qja,Amoros:2013nxa,
Otalora:2013dsa, Geng:2013uga,Nesseris:2013jea, Bamba:2013ooa,
Nashed:2014uta,Harko:2014sja,Harko:2014aja}
\begin{eqnarray}
S=\frac{1}{2\kappa^2}\int d^4 x \,e\,f(T) \, .
\label{teleaction}
\end{eqnarray}
We stress here that although the field equations of TEGR are identical with
those of General Relativity, $f(T)$ modification gives rise to different
equations than $f(R)$ modification.

However, one can use the torsion tensor in order to construct higher-order
torsion invariants, in a similar way that one uses the Riemann tensor in
order to construct higher-order curvature invariants. In particular, in
\cite{Kofinas:2014owa} the invariant
  \begin{eqnarray}
&&T_G=\left(\mathcal{K}^{\kappa}_{\,\,\,\varphi\pi}\mathcal{K}^{\varphi\lambda}_{\,\,\,\,\,
\,\, \rho }\mathcal{K}^{\mu}_{\,\,\,\,\chi\sigma}
\mathcal{K}^{\chi\nu}_{\,\,\,\,\,\,\,\tau}
-2\mathcal{K}^{\kappa\!\lambda}_{\,\,\,\,\,\,\pi}\mathcal{K}^{\mu}_{
\,\,\,\varphi\rho}
\mathcal{K}^{\varphi}_{\,\,\,\chi\sigma}\mathcal{K}^{\chi\nu}_{\,\,\,\,\,\,\tau}\right.
\nn\\
&&\left. \ \ \ \ \  \ \ \ \ \  \
+2\mathcal{K}^{\kappa\!\lambda}_{\,\,\,\,\,\,\pi}\mathcal{K}^{\mu}_{
\,\,\,\,\varphi\rho}
\mathcal{K}^{\varphi\nu}_{\,\,\,\,\,\,\chi}\mathcal{K}^{\chi}_{\,\,\,\,\sigma\tau}
+2\mathcal{K}^{\kappa\!\lambda}_{\,\,\,\,\,\,\pi}\mathcal{K}^{\mu}_{
\,\,\,\,\varphi\rho}
\mathcal{K}^{\varphi\nu}_{\,\,\,\,\,\,\,\sigma,\tau}\right)
\delta^{\pi\rho\sigma\tau}_{\kappa \lambda \mu \nu}
\label{TG}
\end{eqnarray}
was constructed in an arbitrary dimension $D$, where the generalized
$\delta^{\pi\rho\sigma\tau}_{\kappa \lambda \mu \nu}$ is the determinant
of the Kronecker deltas. This invariant is just the Teleparallel Equivalent
of the Gauss-Bonnet combination  $
G=R^{2}-4R_{\mu\nu}R^{\mu\nu}+R_{\mu\nu\kappa\lambda}R^{\mu\nu\kappa\lambda}
$, and in four dimensions it reduces to a topological term. Thus, inspired by
the $f(G)$ extensions
of General Relativity \cite{Nojiri:2005jg,DeFelice:2008wz}, one can consider
general functions $f(T_G)$ in the action too.

Taking the above into account, one can propose a new class of
gravitational modifications as \cite{Kofinas:2014owa}
\begin{eqnarray}
S =\frac{1}{2\kappa^{2}}\!\int d^{4}x\,e\,f(T,T_G)\,,
\label{fGBtelaction}
\end{eqnarray}
which is also valid in higher dimensions. Since $T_G$ is quartic in the torsion tensor, $f(T,T_G)$ gravity is more
general than the $f(T)$ class. Additionally, $f(T,T_G)$ gravity
is obviously different from $f(R,G)$  one
\cite{Nojiri:2005jg,DeFelice:2008wz,Davis:2007ta,
DeFelice:2009aj,Jawad:2013wla}. Note that the usual Einstein-Gauss-Bonnet
theory for $D>4$ arises in the special case
$f(T,T_G)=-T+\alpha T_G$ (with $\alpha$ the Gauss-Bonnet coupling), while
TEGR (that is GR) is obtained for $f(T,T_G)=-T$.

\section{$f(T,T_G)$ cosmology}
\label{fTcosmology}

In order to investigate the cosmological implication of the above action (\ref{fGBtelaction}),
we consider  a spatially flat cosmological ansatz
\begin{equation}
ds^{2}=-dt^{2}+a^{2}(t)\delta_{\hat{i}\hat{j}}dx^{\hat{i}}dx^{\hat{j}}\,,
\label{metriccosmo}
\end{equation}
where $a(t)$ is the scale factor. This metric arises from the diagonal vierbein
\begin{equation}
\label{vierbeincosmo}
e^{a}_{\,\,\,\mu}=\text{diag}(1,a(t),a(t),a(t))
\end{equation}
through (\ref{metrdef}), while the dual vierbein is
$e_{a}^{\,\,\,\mu}=\text{diag}(1,a^{-1}(t),
a^{-1}(t),a^{-1}(t))$, and its determinant $e=a(t)^{3}$. Thus, inserting the
vierbein (\ref{vierbeincosmo}) into relations (\ref{Tquad}) and  (\ref{TG}), we find
\begin{eqnarray}
\label{Tcosmo1}
 &&T=6H^2\\
 &&T_G= 24H^2\big(\dot{H}+H^2\big) ,
 \label{TGcosmo1}
\end{eqnarray}
where $H=\frac{\dot{a}}{a}$ is the Hubble parameter and dots denote
differentiation with respect to $t$.

Finally, in order to acquire a realistic cosmology we additionally consider
a matter action $S_{m}$, corresponding to an  energy-momentum tensor
$\Theta^{\mu\nu}$, focusing on the case of a perfect fluid   of energy
density $\rho_m$ and pressure $p_m$.

As it was showed in \cite{Kofinas:2014owa}, variation of the total action
$S+S_m$ gives in the case of FRW geometry the following Friedmann equations
\begin{equation}
f-12H^{2}f_{T}-T_G f_{T_G}
+24H^{3}\dot{f_{T_G}}=2\kappa^{2}\rho_m
\label{Fr1}
\end{equation}
\begin{equation}
f-4\big(3H^2+\dot{H}\big)f_T-4H\dot{f_T}-T_G
f_{T_G}+\frac{2}{3H}T_G\dot{f_{T_G}}+8H^2\ddot{f_{T_G}}
=-2\kappa^{2} p_m\,,
\label{Fr2}
\end{equation}
where $\dot{f_{T}}=f_{TT}\dot{T}+f_{TT_{G}}\dot{T}_{G}$,
$\dot{f_{T_{G}}}=f_{TT_{G}}\dot{T}+f_{T_{G}T_{G}}\dot{T}_{G}$,
$\ddot{f_{T_{G}}}=f_{TTT_{G}}\dot{T}^{2}+2f_{TT_{G}T_{G}}\dot{T}
\dot{T}_{G}+f_{T_{G}T_{G}T_{G}}\dot{T}_{G}^{\,\,2}+
f_{TT_{G}}\ddot{T}+f_{T_{G}T_{G}}\ddot{T}_{G}$,
with $f_{TT}$, $f_{TT_{G}}$,... denoting multiple partial differentiations
of $f$ with respect to $T$, $T_{G}$. Here, the involved time-derivatives of
$\dot{T}$, $\ddot{T}$, $\dot{T}_{G}$, $\ddot{T}_{G}$ are straightforwardly
obtained using (\ref{Tcosmo1}), (\ref{TGcosmo1}).

Therefore, we can rewrite the Friedmann equations (\ref{Fr1}) and
(\ref{Fr2}) in the usual form
 \begin{eqnarray}
\label{Fr1b}
H^2& =& \frac{\kappa^2}{3}\left(\rho_m + \rho_{DE} \right)   \\
\label{Fr2b}
\dot{H}& =&-\frac{\kappa^2}{2}\left(\rho_m +p_m+\rho_{DE}+p_{DE}\right),
\end{eqnarray}
defining the energy density and pressure of the effective dark energy sector as
\begin{equation}
\label{rhode}
\rho_{DE}\equiv\frac{1}{2\kappa^2}\left( 6H^2
-f+12H^{2}f_{T}+T_G f_{T_G}-24H^{3}\dot{f_{T_G}}\right)
\end{equation}
\begin{equation}
\label{pde}
p_{DE}  \equiv   \frac{1}{2\kappa^2}\left[\!
-2(2\dot{H}\!+\!3H^2)+f\!-\!4\big(\dot{H}\!+\!3H^2\big)f_T-4H\dot{f_T}-T_G
f_{T_G}+\frac{2}{3H}T_G\dot{f_{T_G}}+8H^2\ddot{f_{T_G}}
\right]\!.
\end{equation}
The standard matter $\rho_{m}$ is conserved independently, i.e. $\dot{\rho}_{m}+3H(\rho_{m}+p_{m})=0$.
One can easily verify that the dark energy density and pressure satisfy
the usual evolution equation
\begin{eqnarray}
\dot{\rho}_{DE} +3H(\rho_{DE}+p_{DE})=0,
\end{eqnarray}
and we can also define the dark energy equation-of-state parameter as
usual
\begin{eqnarray}
w_{DE}\equiv \frac{p_{DE}}{\rho_{DE}}.
\end{eqnarray}

\section{Dynamical analysis}
\label{Phasespaceanalysis}

In order to perform the stability analysis of a given cosmological scenario,
one first transforms it to its autonomous form $\label{eomscol}
\textbf{X}'=\textbf{f(X)}$
\cite{Perko,Ellis,Copeland:1997et,Ferreira:1997au,Chen:2008ft,Cotsakis:2013zha,Giambo':2009cc},
where $\textbf{X}$ are some auxiliary variables presented as a
column vector and primes denote derivatives
with respect to $N=\ln a$. Then, one extracts the  critical points
$\bf{X_c}$  by imposing the condition  $\bf{X}'=0$, and in order to determine
their stability properties one expands around them with $\textbf{U}$ the
column vector of the perturbations of the variables. Therefore,
for each critical point the perturbation equations are expanded to first
order as $\label{perturbation} \textbf{U}'={\bf{Q}}\cdot
\textbf{U}$, with the matrix ${\bf {Q}}$ containing the coefficients of the
perturbation equations. The eigenvalues of ${\bf {Q}}$ determine the type and
stability of the specific critical point.

In order to perform the above analysis, we need to specify the  $f(T,T_{G})$
form. In usual $f(T)$ gravity one starts adding corrections of $T$-powers.
However, in the scenario at hand, since $T_G$ contains quartic torsion
terms it is of the same order with $T^2$. Therefore, $T$ and
$\sqrt{T^{2}+\alpha_{2}T_{G}}$
are of the same order, and thus, one should use both in a modified theory.
Hence, the simplest non-trivial model, which does not introduce a new mass
scale  into the problem
and differs from General Relativity, is the one based on
\begin{equation}
f(T,T_G)=-T+\alpha_1\sqrt{T^2+\alpha_2 T_G}\,.
\label{ansantz}
\end{equation}
The couplings $\alpha_{1},\alpha_{2}$ are dimensionless and the model is
expected to play an important role at late times. Indeed, this
model, although simple, can lead to interesting cosmological behavior,
revealing the advantages, the capabilities, and the new features of
$f(T,T_{G})$ cosmology. We mention here that when $\alpha_2=0$ this scenario
reduces to TEGR, that is to General Relativity, with just a rescaled
Newton's constant, whose dynamical analysis has been performed in detail in
the literature \cite{Copeland:1997et,Ferreira:1997au,Chen:2008ft}. Thus, in
the following we restrict our analysis to the case $\alpha_2\neq0$.

In this case, the cosmological equations are the Friedmann equations (\ref{Fr1b}),
(\ref{Fr2b}), with  the effective dark energy density and pressure (\ref{rhode})
and (\ref{pde}) becoming
\begin{equation}
\label{rhodeb}
\!\!\!\!\!\!\!\!\!\!\!\!\!\!\!\!\!\!\!\!\!\!\!\!\!\!\!\!\!\!\!\!\!\!\!\!\!\!
\!\!\!\!\!\!\!\!\!\!\!\!\!\!\!\!\!\!\!\!\!\!\!\!\!\!\!\!\!\!\!\!\!\!\!\!\!\!\!
 \kappa^2 \rho_{DE}=
 \frac{\sqrt{3} \alpha_1 H^2 \left\{\alpha_2^2 \ddot H+9
\alpha_2 H \dot H+\big[(3-2 \alpha_2) \alpha_2+9\big] H^3\right\}}{D^{3/2}}
   \end{equation}
\begin{eqnarray}
\label{pdeb}
&&\!\!\!\!\!\!
\kappa^2 p_{DE}
=\frac{\alpha_1 \!\left\{\!(2
\alpha_2\!+\!3) \big[\alpha_2 (10 \alpha_2\!-\!51)\!-\!18\big] H^4 \!+\!\alpha_2 \big[4
   \alpha_2 (5 \alpha_2\!-\!21)\!-\!90\big] H^2 \dot H \!-\!54 \alpha_2^2 \dot H^2   \right\}\!H\dot{H}}
{\sqrt{3} D^{5/2}}
   \nonumber\\
   &&\ \ \ \ \ \ \ \      -\frac{\alpha_1 \alpha_{2}^{2} H \dddot H}
{\sqrt{3} D^{3/2}}
 +
 \frac{\sqrt{3} \alpha_1 \alpha_2^3 H \ddot H^2}{D^{5/2}}-
\frac{2
   \alpha_1 \alpha_2^2 \ddot H \left[2 (\alpha_2\!-\!3) H^2 \dot H+2 \alpha_2
\dot H^2+(6 \alpha_2\!+\!9) H^4\right]}{\sqrt{3} D^{5/2}}
   \nonumber\\
   &&\ \ \ \ \ \ \ \  + \frac{\sqrt{3}\alpha_1   (\alpha_2\!-\!3) (2
\alpha_2\!+\!3)^2 H^7 }
{ D^{5/2}} \,,
\end{eqnarray}
where $D=3H^2+2\alpha_2(\dot{H}+H^2)$.
In order to perform the dynamical analysis of this cosmological
scenario, we introduce the following auxiliary variables
\begin{eqnarray}
\label{xdefin}
&&x=\sqrt{\frac{D}{3H^2}}=\sqrt{1+\frac{2 \alpha_2}{3}\Big(1+\frac{\dot
H}{H^2}\Big)}\\
&&\Omega_m=\frac{\kappa ^2 \rho_m}{3 H^2}.
\end{eqnarray}
Thus, the cosmological system is transformed to the following autonomous form
\begin{align}
&x'=-\frac{x \left[3 \alpha_1 x^2-6
(1\!-\!\Omega_m)x+\alpha_1 (3\!-\!4 \alpha_2)\right]}{2 \alpha_1 \alpha_2}
\label{eqx}\\
&\Omega_m'= -\frac{\Omega_m \left(3
x^2+\alpha_2+3 \alpha_2 w_m-3\right)}{\alpha_2}\,, \label{eqm}
\end{align}
where primes denote differentiation with respect to the new time variable $N$, so
$f'= H^{-1} \dot f$. The above dynamical system is defined in the phase
space $\left\{(x,\Omega_m)| x\in [0,\infty), \Omega_m\in
[0,\infty]\right\}$.

One can now express the various observables in terms of the above auxiliary
variables $\Omega_m$ and $x$ (note that $\Omega_m$ is an observable
itself, that is the matter density parameter). In particular, the deceleration
parameter $q\equiv -1-\dot{H}/H^2$ is given by
\begin{equation}
q=\frac{3 \left(1-x^2\right)}{2 \alpha_2}.
\label{decc}
\end{equation}
Similarly, the dark energy density parameter straightaway reads
\begin{equation}
\Omega_{DE}\equiv \frac{\kappa^2\rho_{DE}}{3 H^2}= 1-\Omega_m.
\end{equation}
The dark energy equation-of-state parameter $w_{DE}$ is given by the relation
$2q=1+3(w_{m}\Omega_{m}+w_{DE}\Omega_{DE})$, and therefore
\begin{equation}
w_{DE}= \frac{3 x^2+\alpha_2+3 \alpha_2 w_m \Omega_m-3}{3 \alpha_2
(\Omega_m-1)}\,,
\label{wdephasespace}
\end{equation}
where $w_m\equiv \frac{p_m}{\rho_m}$ is the matter equation-of-state
parameter. In the following, without loss of generality we assume dust matter
($w_m=0$), but the extension to general $w_m$ is straightforward.

\subsection{Finite phase space analysis}

We now proceed to the detailed phase-space analysis. The real and
physically interesting (that is corresponding to an expanding
universe) critical points of the
autonomous system \eqref{eqx}-\eqref{eqm}, obtained by setting the
left hand sides of these equations to zero, are presented in Table
\ref{tab1}. In the same table we provide their existence conditions.
Their stability is extracted by examining the sign of the real part
of the eigenvalues of the $2\times2$ matrix ${\bf {Q}}$ of the
corresponding linearized perturbation equations. This procedure is
shown in the Appendix \ref{appendixfinite}, and in Table \ref{tab1}
we summarize the stability results. Furthermore, for each critical
point we calculate the values of the deceleration parameter $q$ and
the dark energy equation-of-state parameter $w_{DE}$ given by
(\ref{decc}) and (\ref{wdephasespace}), and we present the results
in Table \ref{tab2}. Finally, in the same Table we summarize the
physical description of the solutions, which we analyze in the next
section.
\begin{table*}
\resizebox{1.0\textwidth}{!}
{\begin{tabular}
{|c| c| c| c| c| }
 \hline
Cr. P. &$x$&$\Omega_m$ &Existence &Stability\\
\hline\hline
$P_1$  &$\sqrt{1-\frac{\alpha_2}{3}}$ & $\Omega_{m1}$ &  $\frac{6}{5}<\alpha_2<3,  \alpha_1\geq -2\sqrt{\frac{3(3-\alpha_2)}{(-6+
5\alpha_2)^2}}$ or &
Stable spiral for $\alpha_2<3$ and  \\
             &&& $\alpha_2=\frac{6}{5} $ or
&$-32 \sqrt{3}
\sqrt{\frac{(3-\alpha_2)^3}{\left(71 \alpha_2^2-336
\alpha_2+288\right)^2}}<\alpha_1<0$ or  \\
             &&& $\alpha_2<\frac{6}{5}, \, \alpha_1 \leq 2
\sqrt{\frac{3(3-\alpha_2)}{(-6+ 5\alpha_2)^2}}$ & $\alpha_1<0,\alpha_2\leq
\frac{1}{71} \left(168-36 \sqrt{6}\right)\approx 1.124$.\\
&&&& Saddle otherwise
(hyperbolic cases). \\[0.2cm]
\hline
$P_2$ & $x_2$  & $0$ &
$\alpha_2<\frac{3}{4},\,0<\alpha_1\leq \sqrt{\frac{3}{3-4 \alpha_2}}$ or &
Stable node for $\alpha_2<0, \,0<\alpha_1<2 \sqrt{\frac{3(3-\alpha_2)}{(5
\alpha_2-6)^2}}$
   \\
       &&& $ \alpha_1\neq 0, \, \alpha_2=\frac{3}{4}$ or
& or $\frac{6}{5}<\alpha_2\leq
   3, \,\alpha_1<-2 \sqrt{\frac{3(3- \alpha_2)}{(5 \alpha_2-6)^2}}$
\\
             &&& $\alpha_2>\frac{3}{4}, \,\alpha_1<0$ &or
$\alpha_2>3, \,\alpha_1<0$.
\\
             &&&   &Unstable node for $0<\alpha_2<\frac{3}{4},\,
0<\alpha_1<\sqrt{\frac{3}{3-4 \alpha_2}}.$
 \\
             &&&   &Saddle otherwise (hyperbolic cases).
\\[0.2cm]
\hline
$P_3$  & $x_3$ &$0$&
$\alpha_2<\frac{3}{4},\, 0<\alpha_1\leq  \sqrt{\frac{3}{3-4 \alpha_2}}$ or &
 Stable node for
       $\alpha_1>0,\alpha_2\geq \frac{6}{5}.$
   \\
&&& $\alpha_2\geq \frac{3}{4},\,
   \alpha_1>0$ &  Unstable node for  $\alpha_2<0,\,
0<\alpha_1<\frac{\sqrt{3}}{\sqrt{3-4 \alpha_2}}.$
 \\
             &&&   &Saddle otherwise (hyperbolic cases).
\\[0.2cm]
\hline
$P_4$ & $0$ & $0$ & Always & Unstable node for $\frac{3}{4}<\alpha_2<3$.
\\
             &&&   &Saddle otherwise (hyperbolic cases).
\\[0.2cm]
\hline
\end{tabular}}
\caption[crit]{The real and physically interesting
critical points of the autonomous system  \eqref{eqx}-\eqref{eqm}. Existence
and stability conditions. We use the notations
$\Omega_{m1}=\frac{\alpha_1 \sqrt{9-3 \alpha_2}
(6-5 \alpha_2)+6 (\alpha_2-3)}{6
   (\alpha_2-3)}$, $x_{2}=\frac{3-\sqrt{3 \alpha_1^2 (4 \alpha_2-3)+9}}{3
\alpha_1}$ and $x_{3}=\frac{3+\sqrt{3 \alpha_1^2 (4 \alpha_2-3)+9}}{3
\alpha_1}$.}
\label{tab1}
\end{table*}

\begin{table*}
\resizebox{1.0\textwidth}{!}
{\begin{tabular}
{| c|c| c| c| c|}
\hline
Cr. P. &$\Omega_{DE}$ &$q$  & $w_{DE}$ & Properties of solutions\\
\hline\hline
$P_1$ & $1-\Omega_{m1}$  & $\frac{1}{2}$& $0$ & Dark Energy - Dark Matter
scaling solution
\\[0.2cm]
\hline
$P_2$   & $1$ &
$q_2$ & $w_{DE2}$     & Decelerating solution for \\
&&&& $\alpha_2<0,\; \frac{3}{3-2 \alpha_2}<\alpha_1\leq  \sqrt{\frac{3}{3-4 \alpha_2}}$ or  \\
&&&& $0<\alpha_2<\frac{3}{4},\; 0<\alpha_1\leq  \sqrt{\frac{3}{3-4 \alpha_2}}$ or \\
&&&& $\alpha_1\neq 0,\; \alpha_2=\frac{3}{4}$ or\\
&&&& $\frac{3}{4}<\alpha_2\leq \frac{3}{2},\; \alpha_1<0$ or $\alpha_2>\frac{3}{2},\; \frac{3}{3-2 \alpha_2}<\alpha_1<0$.\\
&&&& Quintessence
   solution for \\
&&&&
$\alpha_2\leq -\frac{3}{2},\; \,0<\alpha_1<\frac{3}{3-2 \alpha_2}$ or \\
&&&&$-\frac{3}{2}<\alpha_2<0,\; - \sqrt{\frac{3(2 \alpha_2+3)}{(\alpha_2-3)^2}}<\alpha_1<\frac{3}{3-2 \alpha_2}$ or \\
&&&& $\frac{3}{2}<\alpha_2\leq 3,\; \alpha_1<\frac{3}{3-2 \alpha_2}$ or \\
&&&& $\alpha_2>3,\; - \sqrt{\frac{3(2 \alpha_2+3)}{(\alpha_2-3)^2}}<\alpha_1<\frac{3}{3-2 \alpha_2}$. \\
&&&& De Sitter solution for \\
&&&&
$-\frac{3}{2}<\alpha_2<0,\;\alpha_1=\sqrt{\frac{3(2
\alpha_2+3)}{(\alpha_2-3)^2}}$ or $\alpha_2>3,\;\alpha_1=-\sqrt{\frac{3(2
\alpha_2+3)}{(\alpha_2-3)^2}}$.\\
&&&& Phantom solution for \\
&&&& $-\frac{3}{2}<\alpha_2<0,\;0<\alpha_1<
\sqrt{\frac{3(2\alpha_2+3)}{(\alpha_2-3)^2}}$  or  $\alpha_2>3,\;\alpha_1<-\sqrt{\frac{3(2
\alpha_2+3)}{(\alpha_2-3)^2}}$.
\\[0.2cm]
\hline
$P_3$  & $1$ &
$q_3$ & $w_{DE3}$ & Decelerating solution for  \\
&&&& $\alpha_2<0,\; 0<\alpha_1\leq  \sqrt{\frac{3}{3-4 \alpha_2}}$ or  \\
&&&& $0<\alpha_2<\frac{3}{4},\; \frac{3}{3-2 \alpha_2}<\alpha_1\leq  \sqrt{\frac{3}{3-4 \alpha_2}}$ or \\
&&&&
$\frac{3}{4}\leq \alpha_2<\frac{3}{2},\; \alpha_1>\frac{3}{3-2 \alpha_2}$.\\
&&&&
Quintessence
solution for
\\
&&&& $0<\alpha_2<\frac{3}{2},\;  \sqrt{\frac{3(2 \alpha_2+3)}{(\alpha_2-3)^2}}<\alpha_1<-\frac{3}{2 \alpha_2-3}$ or
\\
&&&&
$\frac{3}{2}\leq \alpha_2<3,\; \alpha_1> \sqrt{\frac{3(2 \alpha_2+3)}{(\alpha_2-3)^2}}$.
\\
&&&&   De Sitter
solution for $0<\alpha_2<3,\; \alpha_1=\sqrt{\frac{3(2\alpha_2+3)}{(\alpha_2-3)^2}}$.\\
&&&& Phantom solution for
\\
&&&& $0<\alpha_2<3,\; 0<\alpha_1<\sqrt{\frac{3(2\alpha_2+3)}{(\alpha_2-3)^2}}$ or $\alpha_2\geq 3,\;\alpha_1>0$.
            \\[0.2cm]
\hline
$P_4$  & $1$ &$\frac{3}{2 \alpha_2}$ &
$\frac{1}{\alpha_2}-\frac{1}{3}$  & Decelerating
solution for $\alpha_2> 0$. \\
&&&& Quintessence DE dominated
solution for $\alpha_2< -\frac{3}{2}$. \\
&&&& De Sitter solution for $\alpha_2=-\frac{3}{2}$.   \\
&&&&  Phantom
solution for
$-\frac{3}{2}<\alpha_2<0$. \\
[0.2cm]
\hline
\end{tabular}}
\caption[crit2]{
The real and physically interesting
critical points of the autonomous system \eqref{eqx}-\eqref{eqm}, and the
corresponding values of the dark energy density parameter $\Omega_{DE}$, the
deceleration parameter $q$ and the dark energy
equation-of-state parameter $w_{DE}$. We use the notation
$\Omega_{m1}=\frac{\alpha_1 \sqrt{9-3 \alpha_2}
(6-5 \alpha_2)+6 (\alpha_2-3)}{6
   (\alpha_2-3)}$,   $q_{2}= \frac{\sqrt{3 \alpha_1^2 (4
\alpha_2-3)+9}}{\alpha_1^2
\alpha_2}-\frac{3}{\alpha_1^2 \alpha_2}+\frac{3}{\alpha_2}-2$,  $q_{3}=-
\frac{\sqrt{3 \alpha_1^2 (4
\alpha_2-3)+9}}{\alpha_1^2
\alpha_2}-\frac{3}{\alpha_1^2 \alpha_2}+\frac{3}{\alpha_2}-2$, $w_{DE
2}= \frac{2
\sqrt{3 \alpha_1^2 (4 \alpha_2-3)+9}}{3 \alpha_1^2
\alpha_2}-\frac{2}{\alpha_1^2
   \alpha_2}+\frac{2}{\alpha_2}-\frac{5}{3}$  and $w_{DE
3}=- \frac{2
\sqrt{3 \alpha_1^2 (4 \alpha_2-3)+9}}{3 \alpha_1^2
\alpha_2}-\frac{2}{\alpha_1^2
   \alpha_2}+\frac{2}{\alpha_2}-\frac{5}{3}$.
In the last column we summarize
their physical description. }
\label{tab2}
\end{table*}

\subsection{Phase space analysis at infinity}

Due to the fact that the dynamical system \eqref{eqx}-\eqref{eqm} is
non-compact, there could be non-trivial dynamical features in the asymptotic
regime too. Therefore, in order to complete the phase space analysis
 we must extend our investigation with the analysis at infinity using the
Poincar\'e projection method \cite{PoincareProj,Xu:2012jf}.

We introduce the new coordinates $(r,\theta)$ defined by
\begin{eqnarray}
&&x=\frac{r}{1-r}\cos\theta\label{Poincvariables1}
\\
&& \Omega_m=\frac{r}{1-r}\sin\theta,
\label{Poincvariables2}
\end{eqnarray}
with $\theta\in
\left[0,\frac{\pi}{2}\right]$ and $r\in \left[0,1\right)$. Thus, the critical
points at infinity, that is $x\rightarrow+\infty$ or
$\Omega_m\rightarrow+\infty$ (that is $R^2\equiv x^2+\Omega_m^2\rightarrow
+\infty$), correspond to $r\rightarrow 1^-$. Moreover, the region of
the plane $(r,\theta)$ that is corresponding
to $0\leq x,\, 0\leq \Omega_m\leq 1$ is given by
\begin{equation}
\label{rest_infty}
 \left\{(r,\theta): 0\leq r\leq \frac{1}{2}, 0\leq \theta\leq
\frac{\pi}{2}\right\}\cup   \left\{(r,\theta): \frac{1}{2}<r<1,
0\leq \theta \leq \arcsin \left(\frac{1-r}{r}\right)\right\}.
\end{equation}
Using relations (\ref{Poincvariables1}), (\ref{Poincvariables2}) and
substituting into (\ref{decc}) and (\ref{wdephasespace}),
we obtain the deceleration and
equation-of-state parameters as a function of the new variables,
namely
\begin{eqnarray}
&&q=\frac{3 \left(1-2r+r^2\sin^2\theta\right)}{2 \alpha_2(1-r)^2}
 \label{q22}\\
&&w_{DE}=\frac{\alpha_2(1-r)^2-3\left(1-2r+r^2\sin^2\theta\right)}{
3\alpha_2(1-r)\left[r(\sin\theta+1)-1\right] }\, ,
\label{wdephase22}
\end{eqnarray}
while $\Omega_{DE}$ is just $1-\Omega_m$, that is
\begin{eqnarray}
 \Omega_{DE}=\frac{1-r(1+\sin\theta)}{1-r}.
 \label{Omegasde22}
\end{eqnarray}
\begin{table*}[!]
\begin{center}
\begin{tabular}{|c|c|c|c|c|c|}
\hline
 Cr. P. & $\theta$ &
Stability & $\Omega_{DE}$ & $q$ &  $w_{DE}$ \\
\hline \hline
$Q_1$& $0$
& saddle point &  $1$ & $-\text{sgn}(\alpha_2) \infty$  &
$-\text{sgn}(\alpha_2) \infty$   \\
\hline
$Q_2$& $\arctan\left(\frac{\alpha_1}{2}\right)$
& unstable for $\alpha_2>0$ &  $-\infty$ &  $-\text{sgn}(\alpha_2) \infty$  &
$\text{sgn}(\alpha_2) \infty$    \\
&& stable for $\alpha_2<0$
&&&\\
\hline
$Q_3$& $\frac{\pi}{2}$
& see numerical elaboration &  $-\infty$ & $\frac{3}{2\alpha_2}$  &
$0$   \\
\hline
\end{tabular}
\end{center}
\caption[crit]{\label{tab3} The real critical points of the
autonomous system \eqref{eqx}-\eqref{eqm} at infinity,
stability conditions, and the
corresponding values of the dark energy density parameter $\Omega_{DE}$, of
the deceleration parameter $q$, and of the dark energy equation-of-state
parameter $w_{DE}$. All points correspond to a form of   future, past, or
intermediate singularity, depending on the parameters
\cite{Sami:2003xv,Nojiri:2005sx,Briscese:2006xu,
Bamba:2008ut,Capozziello:2009hc,
Saridakis:2009jq}.}
\end{table*}

Applying the procedure described in the Appendix \ref{appendixinfin}, we
conclude that there are three critical point at infinity. These critical
points, along with their stability
conditions are presented in Table \ref{tab3}. In the same Table we include
the corresponding values of the observables $\Omega_{DE}$, $q$ and $w_{DE}$,
calculated using (\ref{q22}), (\ref{wdephase22}) and (\ref{Omegasde22}).
These points correspond to Big Rip, sudden or other forms of singularities
\cite{Sami:2003xv,Nojiri:2005sx,Briscese:2006xu,
Bamba:2008ut,Capozziello:2009hc,
Saridakis:2009jq},
depending on whether the singularity is reached at finite or infinite time,
and on their observable features.

\section{Cosmological Implications}
\label{implications}

In the previous section we performed the complete phase-space analysis of
the physically interesting model with $f(T,T_G)=-T+\alpha_1\sqrt{T^2+\alpha_2
T_G}$, both at the finite region and at infinity. Thus, in the present
section we discuss the corresponding cosmological behavior. As usual, the
features of the solutions can be easily deduced by the values of the
observables. In particular, $q<0$ ($q>0$) corresponds to acceleration
(deceleration), $q=-1$ to de Sitter solution, $w_{DE}>-1$ ($w_{DE}<-1$)
corresponds to quintessence-like (phantom-like) behavior, and
$\Omega_{DE}=1$ implies a dark-energy dominated universe.

Point $P_1$ is stable for the conditions presented in Table \ref{tab1}, and
thus it can attract the universe at late times. Since the dark energy and
matter density parameters are of the same order, this point represents a
dark energy - dark matter scaling solution, alleviating the coincidence
problem (note that in order to handle the coincidence problem one should
provide an explanation of why the present $\Omega_m$ and $\Omega_{DE}$ are of
the same order, although they follow different evolution behaviors). However,
it has the disadvantage that
$w_{DE}$ is $0$ and the universe is not accelerating, as expected
\cite{Kofinas:2005hc}.
Although this picture is not favored by observations, it may simply  imply
that the today universe
has not yet reached its asymptotic regime.

Point $P_2$ is stable for the conditions presented in Table \ref{tab1}, and
therefore, it can be the late-times state of the universe. It corresponds to  a
dark energy dominated  universe that can be accelerating. Interestingly
enough, depending on the model
parameters, the dark energy equation-of-state parameter can lie in the
quintessence regime, it can be equal to the cosmological constant value
$-1$, or it can even lie in the phantom regime. These features are  a great
advantage of the scenario at hand,
since they are compatible with observations, and moreover they are obtained
only due to the novel features of $f(T,T_G)$ gravity, without the explicit
inclusion of a cosmological constant or a scalar field, either canonical or
phantom one.

Point $P_3$ is stable for the conditions presented in Table \ref{tab1}, and
therefore, it can attract the universe at late times. It has similar features
with $P_2$, but for different parameter regions. Namely, it corresponds to a
dark energy dominated  universe that can be accelerating, where
the dark energy equation-of-state parameter can lie in the quintessence or
phantom regime, or it can be exactly $-1$. These
features make also this point a good candidate for the description of Nature.

Point $P_4$ corresponds to a dark energy dominated  universe that can be
accelerating, where the dark energy equation-of-state parameter can lie in
the quintessence or phantom regime, or it can be exactly $-1$.  However, $P_{4}$
is not stable and thus it cannot attract the universe at late times.

Finally, the present scenario possesses three critical points at infinity, two
of which can be stable. They correspond to Big Rip, sudden or other forms
of singularities, depending on the parameter choice. We mention that as the
universe moves towards these stable points the matter density parameter
$\Omega_m$ will be larger than $1$. Although this is not theoretically
excluded, growth-index observations could indeed exclude these regions (as it
happens in $f(R)$ gravity \cite{DeFelice:2010aj}), and thus the corresponding
parameter range that
leads the universe to their basin of attraction should be excluded in the
model at hand. Such a detailed investigation has not been performed in
torsion-based gravity, and therefore it has to be done from the beginning.
However, since it lies outside the scope of the present work, it is left for
a future project.

\begin{figure}[ht]
\centering
\subfigure[]{
\includegraphics[scale=0.32]{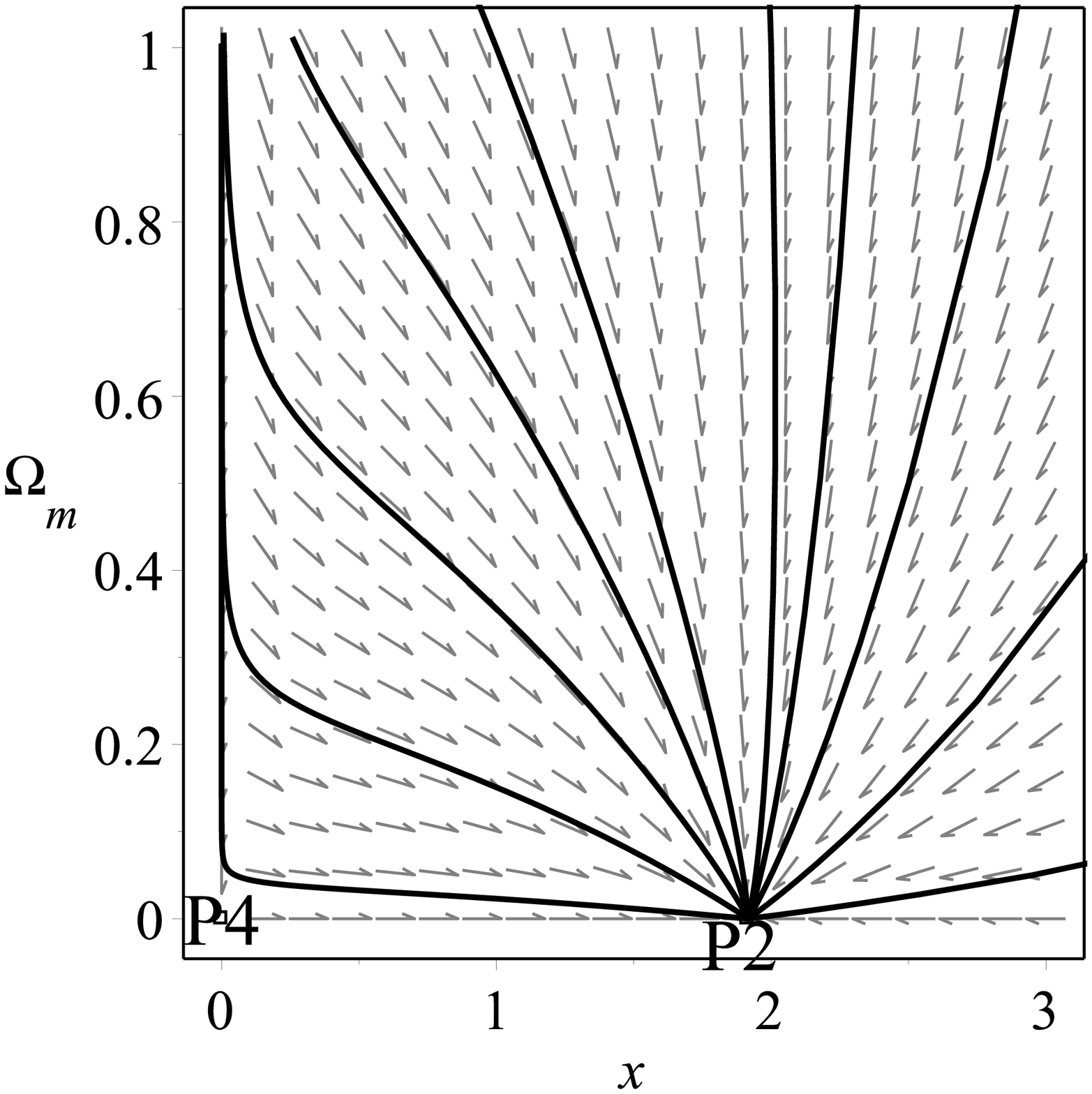}}
\subfigure[]{
\includegraphics[scale=0.30]{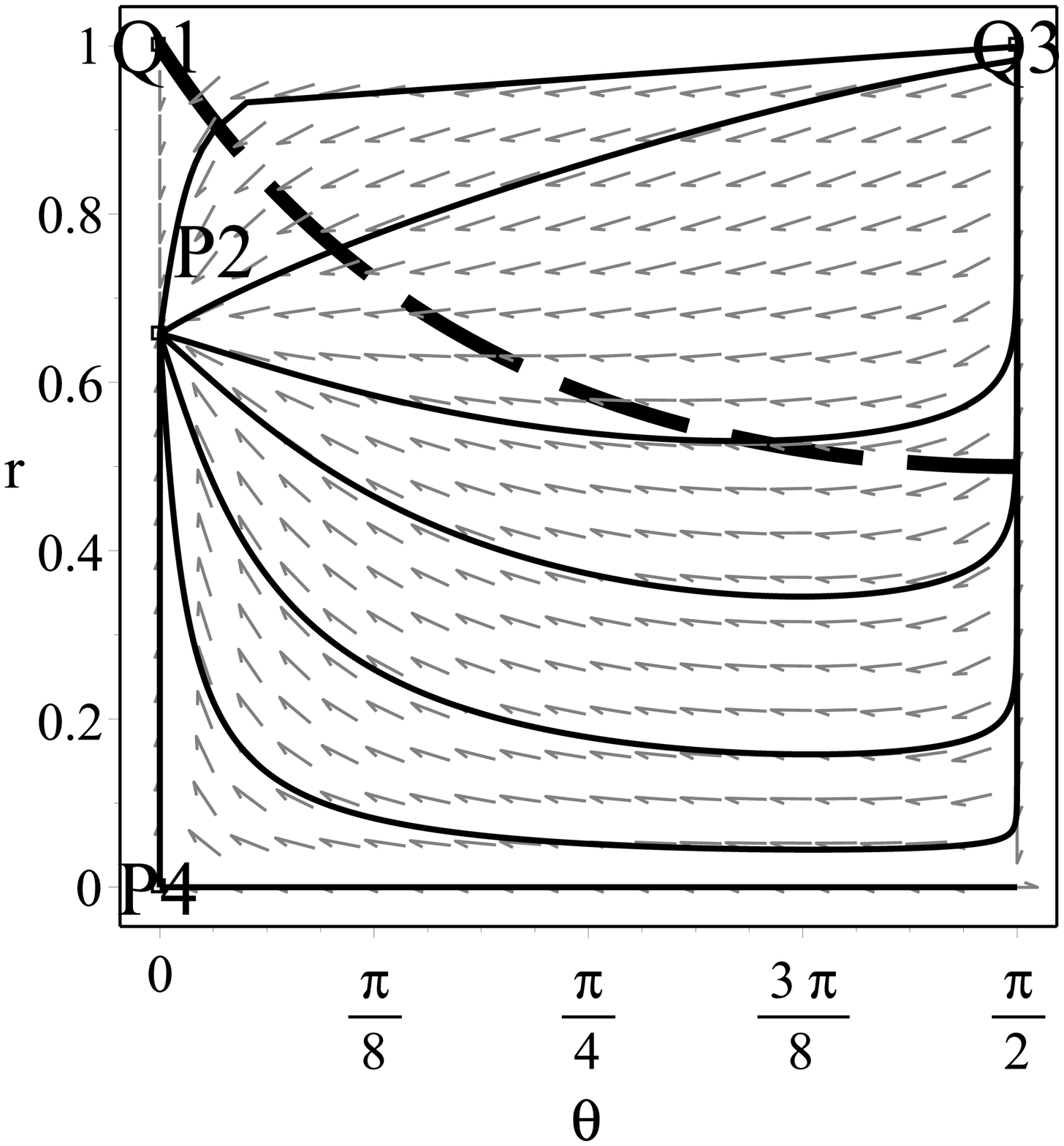}}
\caption{\label{fig1} (a) Trajectories in the  phase space for the
cosmological scenario \eqref{eqx}-\eqref{eqm}, for the parameter choices
 $\alpha_1=-\sqrt{33}$ and $ \alpha_2=4$, and assuming the matter to be dust
($w_m=0$).
(b) Projection of the  phase space on the
``Poincar\'e plane'' $(r,\theta)$. The dashed curve marks the
region above which $\Omega_m>1$ and the universe may result to future
singularities.
 In this specific example the universe is led to the
de Sitter attractor $P_2$, while $P_4$ is saddle. At infinity, there is not
any stable
point, and thus the universe cannot result in any form of singularity
($Q_1$ and $Q_2$ are saddle points ($Q_2$ has $\theta<0$ and thus it is not
depicted in the plot), while $Q_3$ is unstable). }
\end{figure}

In order to present the aforementioned behavior more
transparently, we first evolve the autonomous system \eqref{eqx}-\eqref{eqm}
numerically for the parameter choices $\alpha_1=-\sqrt{33}$ and $
\alpha_2=4$, assuming the matter to be dust ($w_m=0$). The corresponding
phase-space behavior is depicted in Fig. \ref{fig1}. For
completeness we also present the projection in the ``Poincar\'e plane''
$(r,\theta)$, where we depict the behavior at both the finite
and the infinite region. In this case the universe at late times is
attracted by the dark-energy dominated de Sitter attractor $P_2$, where the
effective dark energy behaves like a cosmological constant. At infinity,
there is not any stable point, and thus the universe cannot result in any
form of singularity.
\begin{figure}[t]
\centering
\subfigure[]{
\includegraphics[scale=0.32]{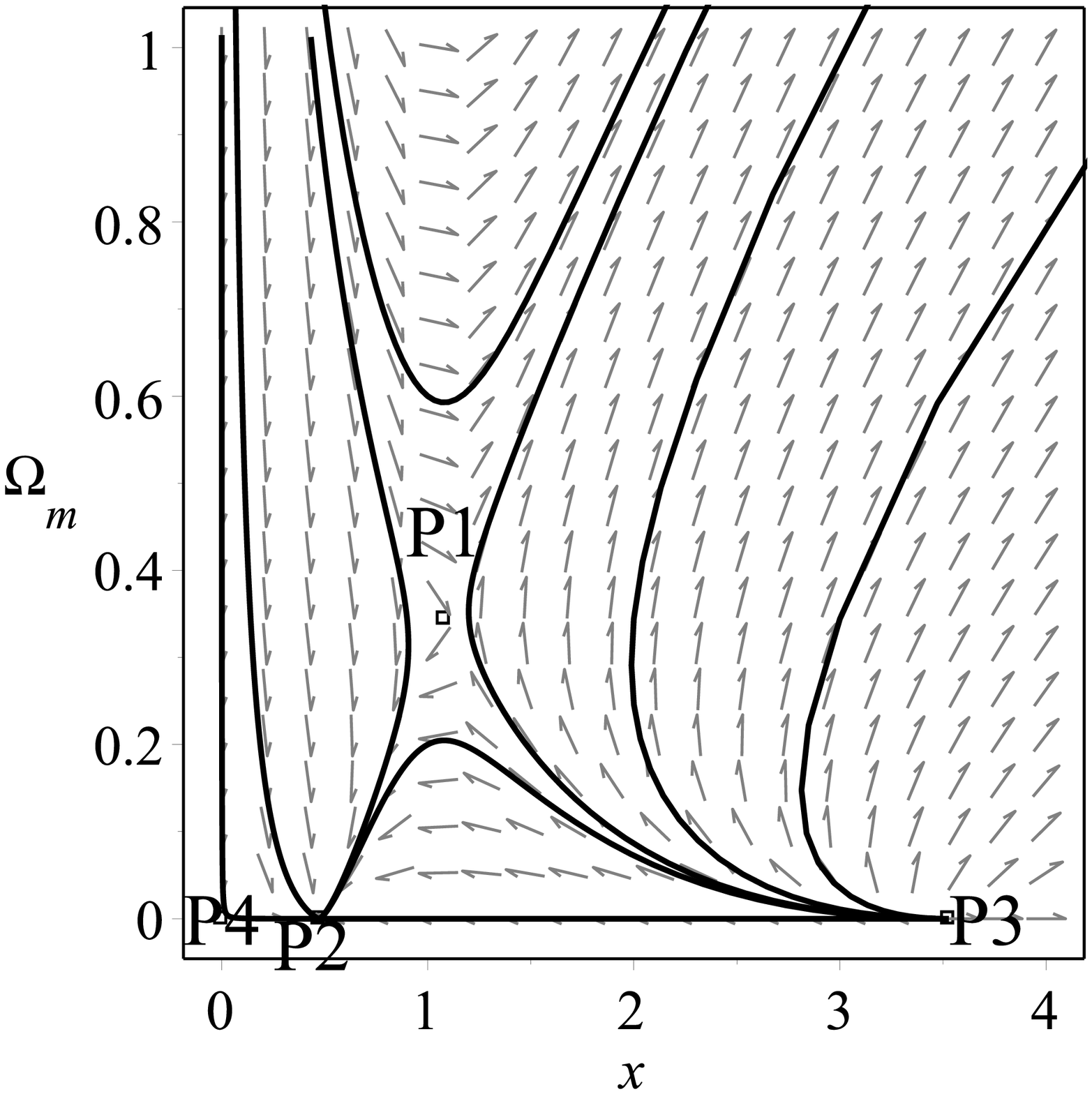}}
\subfigure[]{
\includegraphics[scale=0.30]{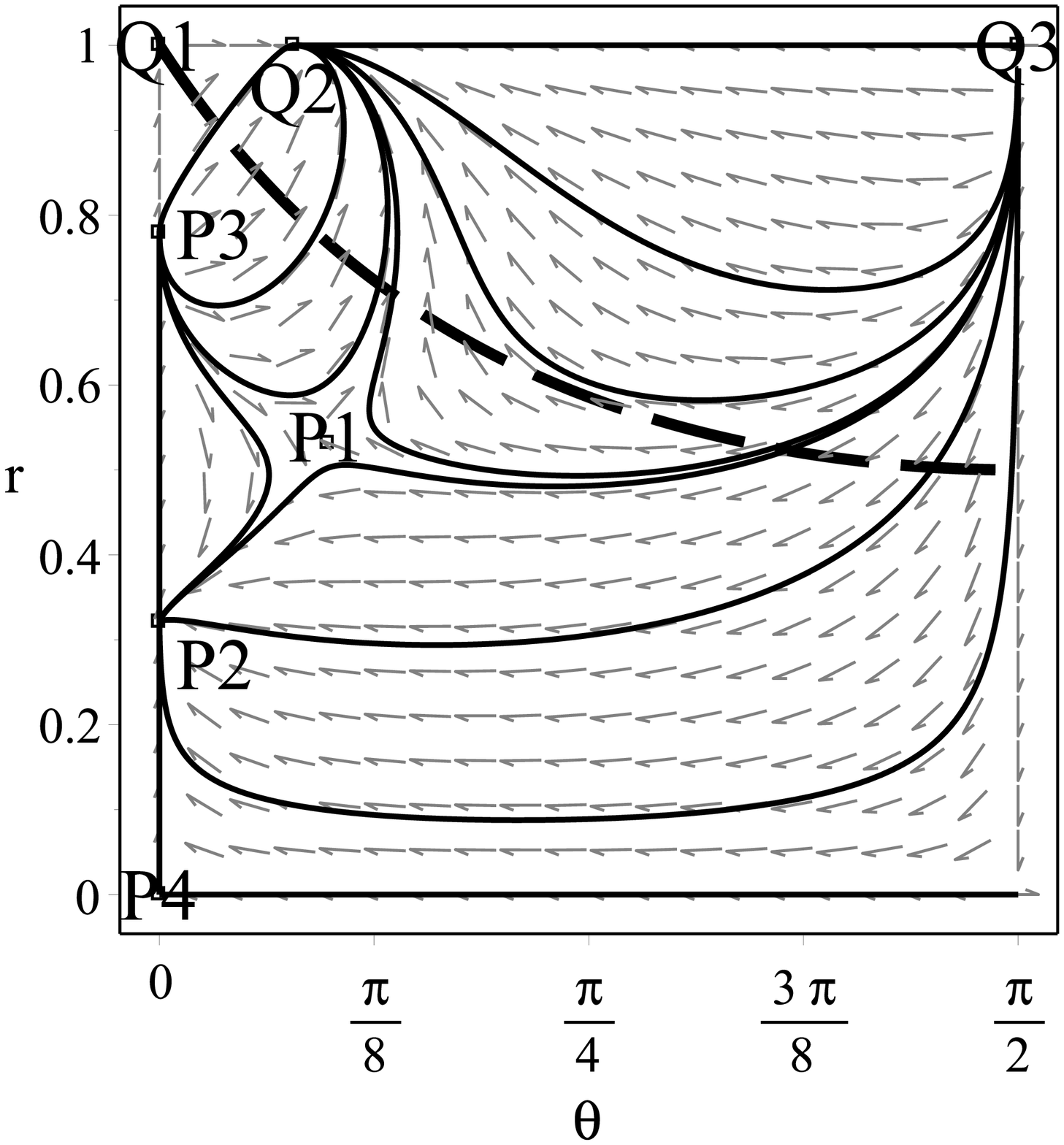}}
\caption{\label{fig2}
(a) Trajectories in the  phase space for the
cosmological scenario \eqref{eqx}-\eqref{eqm}, for the parameter choices
 $\alpha_1=\frac{1}{2}$ and $ \alpha_2=-\frac{1}{2}$, and assuming the matter
to be dust
($w_m=0$).
(b) Projection of the  phase space on the
``Poincar\'e plane'' $(r,\theta)$. The dashed curve marks the
region above which $\Omega_m>1$ and the universe may result to future
singularities.
 In this specific example the universe is led to the
phantom solution $P_2$. At infinity, there is the stable point $Q_2$,
which corresponds to a future singularity.
}
\end{figure}
\begin{figure}[t]
\centering
\subfigure[]{
\includegraphics[scale=0.32]{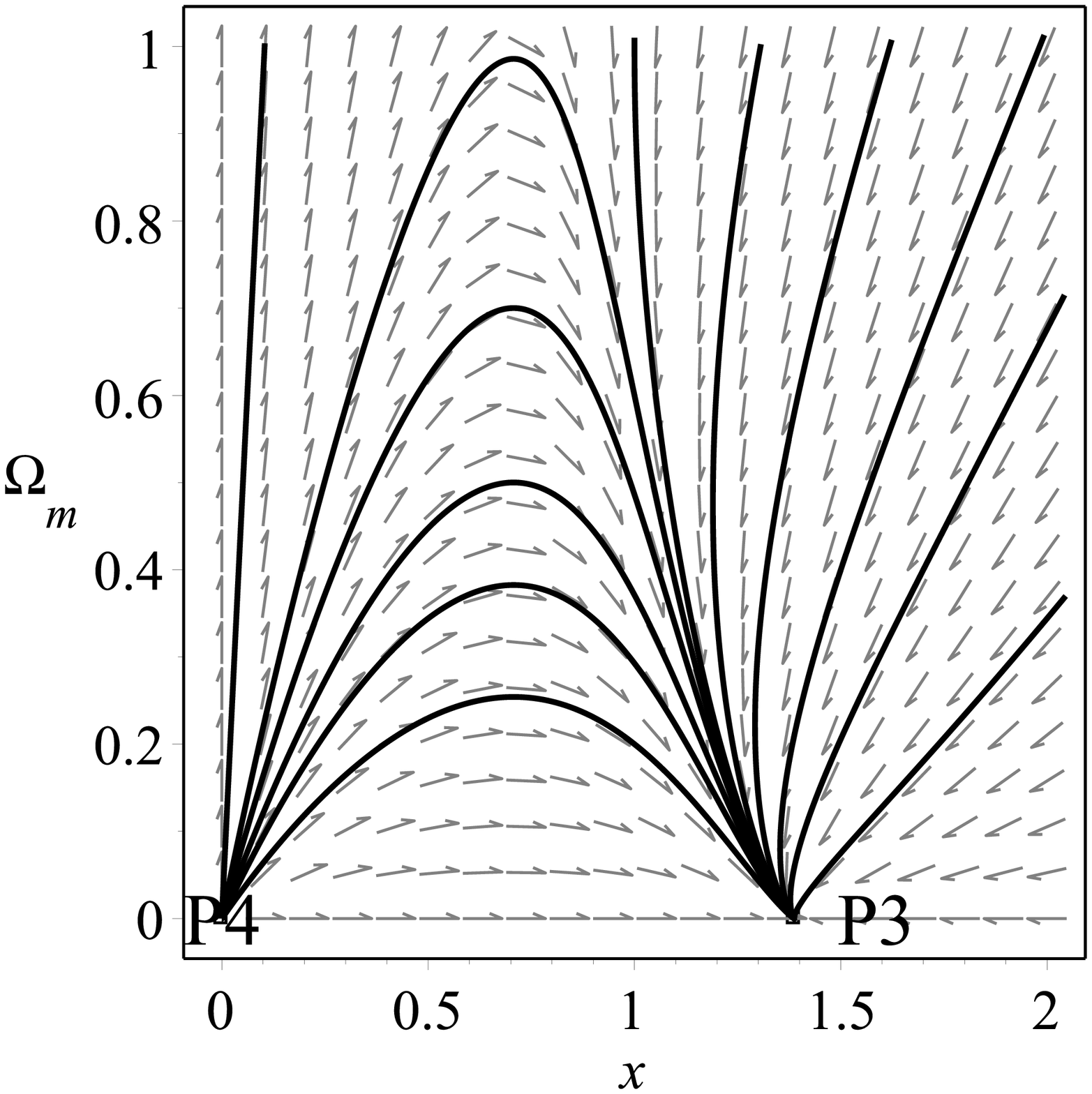}}
\subfigure[]{
\includegraphics[scale=0.30]{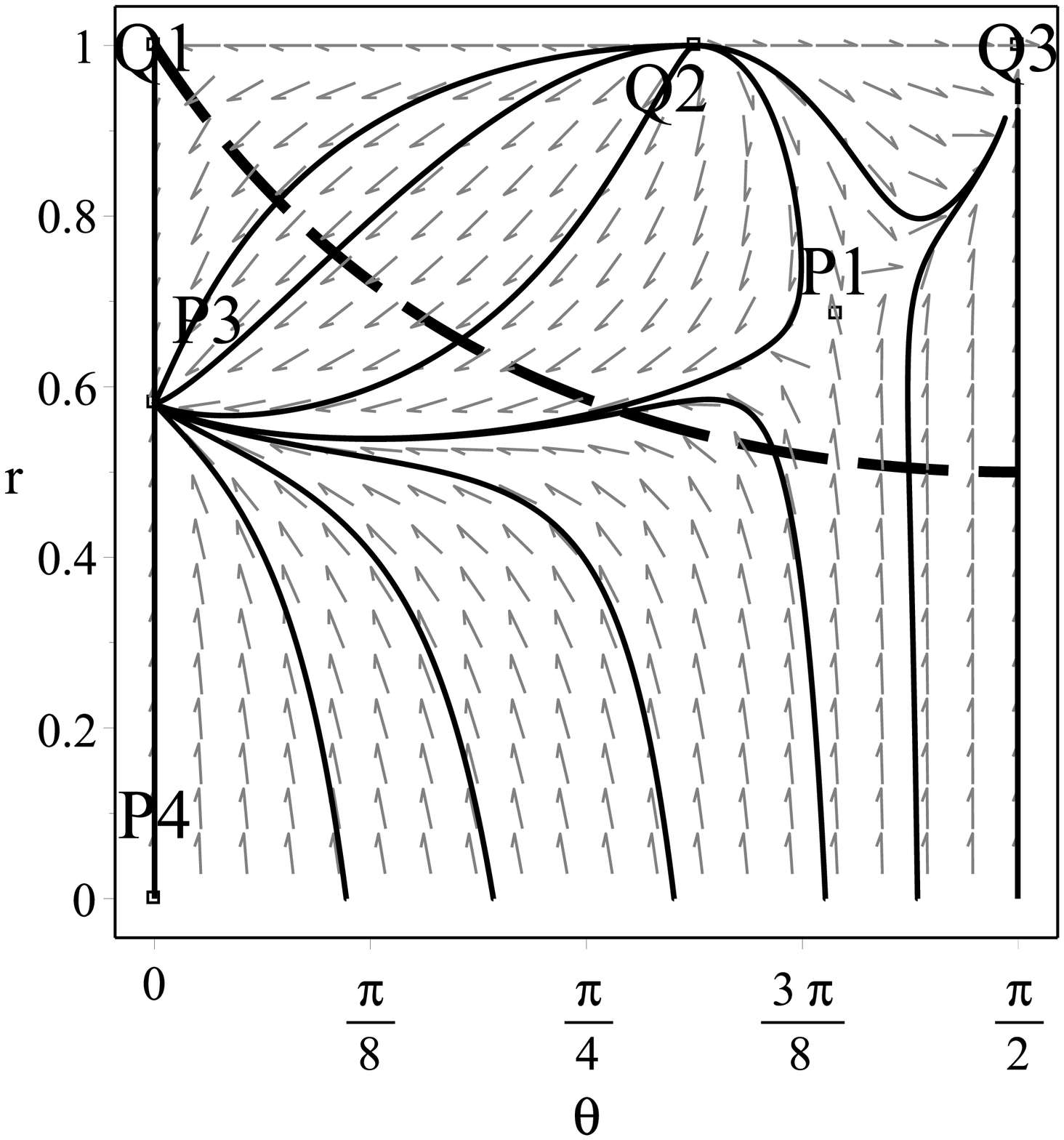}}
\caption{\label{fig5}
(a) Trajectories in the  phase space for the
cosmological scenario \eqref{eqx}-\eqref{eqm}, for the parameter choices
 $\alpha_1=3$ and $ \alpha_2=\frac{3}{2}$, and assuming the matter to be dust
($w_m=0$).
(b) Projection of the  phase space on the
``Poincar\'e plane'' $(r,\theta)$. The dashed curve marks the
region above which $\Omega_m>1$ and the universe may result to future
singularities.
 In this specific example the universe is led to
the  quintessence solution $P_3$. At infinity, there is the stable point
$Q_3$,
which corresponds to a future singularity.
}
\end{figure}

In  Fig. \ref{fig2} we present the phase-space behavior of the autonomous
system \eqref{eqx}-\eqref{eqm} for the choice $\alpha_1=\frac{1}{2}$ and $
\alpha_2=-\frac{1}{2}$ (assuming   $w_m=0$), and its projection on
the ``Poincar\'e plane'' $(r,\theta)$. In this case the attractor at the
finite region is the  phantom solution $P_2$. Additionally, the attractor
in the infinite region is $Q_2$, that is a future singularity.

Finally, in Fig. \ref{fig5} are present some orbits  and the corresponding
Poincar\'e projections for the choice   $\alpha_1=3$ and $
\alpha_2=\frac{3}{2}$, with $w_m=0$. In this case, the universe is attracted
by the quintessence solution $P_3$. Furthermore, in the infinite region the
attractor is $Q_3$, that is a future singularity. \footnote{ Figures
\ref{fig2} and \ref{fig5} are suggestive that the stable manifold of $P_1$
acts as a separatrix, separating the phase-space solutions which are very
likely to end at a future singularity ``at infinity'' from those ending at
the finite region. Hence, the detailed examination of the stable manifold of
$P_1$ may give information of the basin of attraction of the future
singularities.}

\section{Conclusions} \label{Conclusions}

In the present work we studied the dynamical behavior of the recently
proposed scenario of $f(T,T_G)$ cosmology \cite{Kofinas:2014owa}. This class
of modified gravity is based on the quadratic torsion scalar $T$, which is
the Lagrangian of the teleparallel equivalent of General Relativity, as well
as on the new quartic torsion scalar $T_G$, which is the teleparallel
equivalent of the Gauss-Bonnet term. Obviously, $f(T,T_G)$ theories are more
general and cannot be spanned by the simple $f(T)$ ones, and additionally
they are different from $f(R,G)$ class of curvature modified gravity too.

Without loss of generality, as a simple, but non-trivial example, capable of
revealing the advantages and the new features of the theory, we considered a
model where $T$ and $T_G$ corrections are of the same order, and thus
expected to play an important role at late times. We performed for a
spatially flat universe the complete
and detailed phase-space behavior, both in the finite and infinite regions,
calculating additionally also the values of basic observables such is the
various density parameters, the deceleration parameter and the dark energy
equation-of-state parameter.

This scenario exhibits interesting cosmological behaviors. In particular,
depending on the model parameters, the
universe can result in a dark energy dominated accelerating solution and the
dark energy equation-of-state parameter can lie in the quintessence regime,
it can be equal to the cosmological constant value $-1$, or it can even lie in
the phantom regime. Additionally, it can
result in a dark energy - dark matter scaling solution, and thus it can
alleviate the coincidence problem. Finally, under certain parameter choices
the universe can result to Big Rip, sudden, or other form of singularities,
as it is usual in many modified gravitational theories. Definitely, before
the scenario at hand can be considered as a good candidate for the
description of Nature, a detailed confrontation with observations should be
performed. In particular, one should use data from local gravity experiments
(Solar System observations), as well as type Ia
Supernovae (SNIa), Baryon Acoustic Oscillations (BAO), and Cosmic Microwave
Background (CMB) radiation data, in order to impose constraints on the
model. These necessary investigations lie beyond the scope of the present
work and are left for a future project.

\begin{acknowledgments}
GL was supported by COMISI\'ON NACIONAL DE CIENCIAS Y
TECNOLOG\'IA through Proyecto FONDECYT DE POSTDOCTORADO 2014  grant  3140244
and by DI-PUCV grant 123.730/2013.
The research of ENS is implemented within the framework of the Action
``Supporting Postdoctoral Researchers'' of the Operational Program
``Education and Lifelong Learning'' (Actions Beneficiary: General Secretariat
for Research and Technology), and is co-financed by the European Social Fund
(ESF) and the Greek State.
\end{acknowledgments}

\begin{appendix}

\section{Stability of the finite critical points}
\label{appendixfinite}

For the critical points $(x_c,\Omega_{mc})$ of the autonomous system
\eqref{eqx}-\eqref{eqm}, presented in Table \ref{tab1}, the coefficients of
the perturbation equations form a
$2\times2$ matrix ${\bf {Q}}$, which reads:
\begin{eqnarray}
{\bf
{Q}}_{11}&=&\frac{\alpha_1 (4 \alpha_2-3)-9 \alpha_1 x^2-12 x (\Omega_m-1)}{2 \alpha_1 \alpha_2}
\nonumber\\
{\bf
{Q}}_{12}&=&-\frac{3 x^2}{\alpha_1 \alpha_2}
\nonumber\\
{\bf
{Q}}_{21}&=&-\frac{6 x \Omega_m}{\alpha_2}
\nonumber\\
{\bf
{Q}}_{22}&=&-\frac{\alpha_2+3 x^2-3}{\alpha_2}
\nonumber
\end{eqnarray}
Thus, we can straightforwardly see that using
the explicit critical points shown in Table \ref{tab1}, the matrix
${\bf {Q}}$ acquires a simple form that allows for an easy calculation of
its eigenvalues. Hence, by examining the sign of the real parts of
these eigenvalues, we can classify the corresponding critical point. In
particular, if all eigenvalues of a critical point have positive
real parts  then this point is unstable, if they all have negative
real parts then it is stable, and if they change sign then it is a
saddle one. In the following we present the results for each separate
point.

Point $P_1$ has the coordinates
$$P_1: (x,\Omega_m)=\left( \sqrt{1-\frac{\alpha_2}{3}}, \frac{\alpha_1
\sqrt{9-3 \alpha_2} (6-5 \alpha_2)+6 (\alpha_2-3)}{6
   (\alpha_2-3)}\right),$$
   that is it exists for either $
\alpha_2=\frac{6}{5}$ or
$\frac{6}{5}<\alpha_2<3,\, \alpha_1 \geq -2\sqrt{\frac{3(3-\alpha_2)}{(5\alpha_2-6)^2}}$ or
$\alpha_2<\frac{5}{6},\, \alpha_1 \leq 2
\sqrt{\frac{3(3-\alpha_2)}{(5\alpha_2-6)^2}}$.
The eigenvalues of the corresponding linearization matrix are
\begin{align}
\left\{-\frac{\sqrt{\alpha_1 \left[(336-71 \alpha_2)
\alpha_2-288\right]+32
\sqrt{3} (3-\alpha_2)^{3/2}}}{4 \sqrt{\alpha_1} \alpha_2}-\frac{3}{4},
\right. \nonumber \\ \left.  \frac{\sqrt{\alpha_1
\left[(336-71 \alpha_2) \alpha_2-288\right]+32 \sqrt{3}
(3-\alpha_2)^{3/2}}}{4
\sqrt{\alpha_1}
   \alpha_2}-\frac{3}{4}\right\}.
    \end{align}
    Therefore, $P_1$ is a stable spiral for $$
\alpha_2<3,\ \  -32 \sqrt{3}
\sqrt{\frac{(3-\alpha_2)^3}{\left(71 \alpha_2^2-336
\alpha_2+288\right)^2}}<\alpha_1<0,$$ or  $$\alpha_1<0,\alpha_2\leq \frac{1}{71} \left(168-36 \sqrt{6}\right)\lesssim 1.12421.$$ Otherwise it is a saddle (we have excluded the parameter values
that leads to non-hyperbolic critical points).

Point $P_2$ has the coordinates
 $$P_2: (x,\Omega_m)=\left(\frac{3-\sqrt{3 \alpha_1^2 (4
\alpha_2-3)+9}}{3 \alpha_1}, 0\right),$$
that is it exists for
either
$\alpha_2<\frac{3}{4},0<\alpha_1\leq \sqrt{\frac{3}{3-4 \alpha_2}}$ or
$  \alpha_1\neq 0,  \alpha_2=\frac{3}{4}$ or
$
   \alpha_2>\frac{3}{4}, \alpha_1<0$.
 The eigenvalues of the linearization matrix are
    \begin{equation}\!\!\!
    \left\{\frac{2 \sqrt{3 \alpha_1^2 (4 \alpha_2-3)+9}}{\alpha_1^2
\alpha_2}-\frac{6}{\alpha_1^2 \alpha_2}+\frac{6}{\alpha_2}-5, \,
\frac{\sqrt{3 \alpha_1^2 (4 \alpha_2-3)+9}}{\alpha_1^2
\alpha_2}-\frac{3}{\alpha_1^2 \alpha_2}+\frac{3}{\alpha_2}-4\right\}.
    \end{equation}
    Hence, it is an stable node for either
    $$\alpha_2<0, \  \ 0<\alpha_1<2
\sqrt{\frac{3(3-\alpha_2)}{(5
\alpha_2-6)^2}}$$ or
$$\frac{6}{5}<\alpha_2\leq
   3,\  \  \alpha_1<-2 \sqrt{\frac{3(3- \alpha_2)}{(5 \alpha_2-6)^2}}$$ or
$$\alpha_2>3,\  \ \alpha_1<0.$$
Additionally, it is unstable node for
$$0<\alpha_2<\frac{3}{4},\  \
0<\alpha_1<\sqrt{\frac{3}{3-4 \alpha_2}}.$$ Finally, for the remaining
parameter range in the hyperbolic domain, the point behaves as a saddle.

Point $P_3$ has the coordinates
  $$P_3: (x, \Omega_m)=\left( \frac{3+\sqrt{3 \alpha_1^2 (4
\alpha_2-3)+9}}{3 \alpha_1},  0\right),$$ that is it exists for
$\alpha_2<\frac{3}{4},\, 0<\alpha_1\leq  \sqrt{\frac{3}{3-4 \alpha_2}}$ or
$\alpha_2\geq \frac{3}{4},
   \alpha_1>0$.
    The eigenvalues of the corresponding linearization matrix are
\begin{align}
\left\{-\frac{2 \sqrt{3} \sqrt{4 \alpha_1^2 \alpha_2-3
\alpha_1^2+3}}{\alpha_1^2 \alpha_2}-\frac{6}{\alpha_1^2
   \alpha_2}+\frac{6}{\alpha_2}-5, \,
-\frac{\sqrt{3} \sqrt{4 \alpha_1^2 \alpha_2-3 \alpha_1^2+3}}{\alpha_1^2
   \alpha_2}-\frac{3}{\alpha_1^2 \alpha_2}+\frac{3}{\alpha_2}-4\right\}.
\end{align}
That is, it is stable node for $$\alpha_1>0,\ \  \alpha_2\geq \frac{6}{5},$$ it is
unstable node for
$$\alpha_2<0, \ \ 0<\alpha_1<\frac{\sqrt{3}}{\sqrt{3-4 \alpha_2}},$$
otherwise it is a saddle with the exclusion of the parameter values that leads to non-hyperbolic critical point.

Point $P_4$ has the coordinates  $P_4: (x,\Omega_m)=(0,0)$ and it exists
always. The eigenvalues of the linearization matrix read
$$\left\{2-\frac{3}{2
\alpha_2},\,\frac{3}{\alpha_2}-1\right\}.$$
Therefore, it is unstable node for
$\frac{3}{4}<\alpha_2<3$, it is non-hyperbolic for $\alpha_2\in \left\{\frac{3}{4},3\right\}$, otherwise it is a saddle.

The above results are summarized in Table \ref{tab4}.

\begin{table*}
\resizebox{1.0\textwidth}{!}
{\begin{tabular}
{|c| c| c| c| c|c| }
 \hline
Cr. P. &$x$&$\Omega_m$ &Existence & $\nu_1$  & $\nu_2$  \\
\hline\hline
$P_1$  &$\sqrt{1-\frac{\alpha_2}{3}}$ & $\Omega_{m1}$ &  $
\alpha_2=\frac{6}{5}$ or &
$-\frac{\sqrt{\alpha_1 \left[(336-71 \alpha_2)
\alpha_2-288\right]+32
\sqrt{3} (3-\alpha_2)^{3/2}}}{4 \sqrt{\alpha_1} \alpha_2}-\frac{3}{4}$     &
$\frac{\sqrt{\alpha_1
\left[(336-71 \alpha_2) \alpha_2-288\right]+32 \sqrt{3}
(3-\alpha_2)^{3/2}}}{4
\sqrt{\alpha_1}
   \alpha_2}-\frac{3}{4}$ \\
         &&& $\frac{6}{5}<\alpha_2<3, -2\sqrt{\frac{3(3-\alpha_2)}{(-6+
5\alpha_2)^2}}\leq \alpha_1\leq 0$
&  &  \\
             &&&or $\alpha_2=\frac{6}{5} $
& & \\
             &&& or $\alpha_2<\frac{6}{5}, 0\leq \alpha_1 \leq 2
\sqrt{\frac{3(3-\alpha_2)}{(-6+ 5\alpha_2)^2}}$ & &
\\[0.2cm]
\hline
$P_2$ & $\frac{3-\sqrt{3 \alpha_1^2 (4 \alpha_2-3)+9}}{3 \alpha_1}$  & $0$ &
$\alpha_2<\frac{3}{4},0<\alpha_1\leq \sqrt{\frac{3}{3-4 \alpha_2}}$ or &
$\frac{2 \sqrt{3 \alpha_1^2 (4 \alpha_2-3)+9}}{\alpha_1^2
\alpha_2}-\frac{6}{\alpha_1^2 \alpha_2}+\frac{6}{\alpha_2}-5$   &
$\frac{\sqrt{3 \alpha_1^2 (4 \alpha_2-3)+9}}{\alpha_1^2
\alpha_2}-\frac{3}{\alpha_1^2 \alpha_2}+\frac{3}{\alpha_2}-4$
   \\
       &&& $ \alpha_1\neq 0,  \alpha_2=\frac{3}{4}$ or
& &
\\
             &&& $\alpha_2>\frac{3}{4}, \alpha_1<0$ &&
\\[0.2cm]
\hline
$P_3$  & $\frac{3+\sqrt{3 \alpha_1^2 (4 \alpha_2-3)+9}}{3 \alpha_1}$ &$0$&
$\alpha_2<\frac{3}{4}, 0<\alpha_1\leq  \sqrt{\frac{3}{3-4 \alpha_2}}$ or &
$-\frac{2 \sqrt{3} \sqrt{4 \alpha_1^2 \alpha_2-3
\alpha_1^2+3}}{\alpha_1^2 \alpha_2}-\frac{6}{\alpha_1^2
   \alpha_2}+\frac{6}{\alpha_2}-5$ &
$-\frac{\sqrt{3} \sqrt{4 \alpha_1^2 \alpha_2-3 \alpha_1^2+3}}{\alpha_1^2
   \alpha_2}-\frac{3}{\alpha_1^2 \alpha_2}+\frac{3}{\alpha_2}-4$
   \\
&&& $\alpha_2\geq \frac{3}{4},
   \alpha_1>0$ & &
\\[0.2cm]
\hline
$P_4$ & $0$ & $0$ & Always & $2-\frac{3}{2
\alpha_2}$  &
$\frac{3}{\alpha_2}-1$
\\[0.2cm]
\hline
\end{tabular}}
\caption[crit]{The real and physically interesting
critical points at the finite region of the autonomous system  \eqref{eqx}-\eqref{eqm}, their
existence conditions, and the corresponding eigenvalues $\nu_{1},\nu_{2}$ of the matrix
${\bf {Q}}$ of the perturbation equations. We denote
$\Omega_{m1}=\frac{\alpha_1 \sqrt{9-3 \alpha_2}
(6-5 \alpha_2)+6 (\alpha_2-3)}{6
   (\alpha_2-3)}$.}
\label{tab4}
\end{table*}

\section{Stability of the critical points at infinity}
\label{appendixinfin}

We introduce the new coordinates $(r,\theta)$ defined by
\begin{eqnarray}
&&x=\frac{r}{1-r}\cos\theta\nonumber\\
&& \Omega_m=\frac{r}{1-r}\sin\theta,
\end{eqnarray}
with $\theta\in
\left[0,\frac{\pi}{2}\right]$ and $r\in \left[0,1\right)$.
The limit $r\rightarrow 1^-$ corresponds to
$R^2\equiv x^2+\Omega_m^2\rightarrow
\infty.$ Note that the physical region of the plane $(r,\theta)$, that is
corresponding
to $0\leq x,\, 0\leq \Omega_m\leq 1$, is given by
\begin{equation}
\label{rest_infty2}
 \left\{(r,\theta): 0\leq r\leq \frac{1}{2}, 0\leq \theta\leq
\frac{\pi}{2}\right\}\cup   \left\{(r,\theta): \frac{1}{2}<r<1,
0\leq \theta \leq \arcsin \left(\frac{1-r}{r}\right)\right\}.
\end{equation}
The leading terms of the equations for $r'$ and $\theta'$ as $r\rightarrow
1^-$ are
\begin{eqnarray}
&&r' \rightarrow \frac{3 \cos ^2(\theta ) [\alpha_1 (\cos (2 \theta )-3)-2
\sin
(2 \theta )]}{4 \alpha_1 \alpha_2 (1-r)}
\label{inftyr}\\
&&\theta'\rightarrow -\frac{3 \sin (\theta ) \cos ^2(\theta ) [\alpha_1 \cos
(\theta )-2 \sin (\theta )]}{2 \alpha_1 \alpha_2 (1-r)^2}
\label{inftytheta}.
\end{eqnarray}
Hence, the fixed points at infinity (that is for $r\rightarrow 1^-$) are
obtained by setting $\theta'=0$, and solving for $\theta$.

Let us denote a generic fixed
point by $\theta=\theta^*$. The stability of this point is studied by
analyzing first the stability of the angular coordinates from equation
\eqref{inftytheta}, and then deducing, from the sign of the equation
\eqref{inftyr}, the stability on the radial direction \footnote{The special
functional form of the terms in the denominator depending on $r$ is
irrelevant
for the discussion, since they can be removed by choosing a different time
scale. What is important is that the sign of these terms is positive, which
implies that the arrow of time is preserved under the time rescaling.}.
 A fixed point $\theta=\theta^*$ is said to be stable if both
\begin{equation}
\frac{d \theta'}{d\theta}\left|_{\theta=\theta^*}\right. <0, r'
\left|_{\theta=\theta^*} \right. >0.
\end{equation}
The first condition implies stability of the angular coordinate $\theta$. The
second condition implies that the $r$-values increase before reaching
the
limit value $r=1$ (that is before the boundary ``at infinity'' is reached) at
the fixed point $\theta=\theta^*$.
Similarly, a fixed point $\theta=\theta^*$ is said to be unstable if both
\begin{equation}
\frac{d \theta'}{d\theta}\left|_{\theta=\theta^*}\right. >0, r'
\left|_{\theta=\theta^*}\right. <0.
\end{equation}
Finally, it is a saddle point if either
\begin{equation}
\frac{d \theta'}{d\theta}\left|_{\theta=\theta^*}\right. >0, r'
\left|_{\theta=\theta^*} \right. >0,
\end{equation}
or
\begin{equation}
\frac{d \theta'}{d\theta}\left|_{\theta=\theta^*}\right. <0, r'
\left|_{\theta=\theta^*} \right. <0.
\end{equation}

In summary, the fixed points of the autonomous system at hand at infinity are
the following:

\begin{itemize}
\item
$Q_1: \theta^*=0, r^*=1, x=\infty, \Omega_m=0$.

Since from the definition (\ref{xdefin}) we have $x=\sqrt{1+\frac{2
\alpha_2}{3}\left(1+\frac{\dot
H}{H^2}\right)}$, we deduce that the corresponding cosmological
solution satisfies
$\text{sign}(\alpha_2\dot H)\frac{|\dot H|}{H^2}\rightarrow \infty,$ or
$H\rightarrow 0$ ($\dot H$ is bounded, with $\text{sign}(\alpha_2 \dot H)
>0$). Since $q=w_{DE}=-\text{sgn}(\alpha_2)\infty$, the  point represents a
super-accelerated phantom solution for $\alpha_2>0$, where eventually the
universe ends in  Big Rip, sudden or other forms of singularities
(depending on whether the singularity is reached at finite or infinite time,
what are its features etc.)
\cite{Sami:2003xv,Nojiri:2005sx,Briscese:2006xu,
Bamba:2008ut,Capozziello:2009hc,
Saridakis:2009jq}.
For
$\alpha_2<0$ it is a decelerating solution where the universe
asymptotically stops expanding. Since $\left(\frac{d \theta'}{d
\theta},r'\right)|_{\theta=0}=\left(-\frac{3}{2\alpha_2},-\frac{3}{2\alpha_2}
\right)$, we conclude that  $Q_1$ is always a saddle point.

\item
$Q_2: \theta^*=\arctan\left(\frac{\alpha_1}{2}\right), \alpha_1\neq 0, r^*=1,
\frac{\Omega_m}{x}\rightarrow \frac{\alpha_1}{2}, x\rightarrow \infty,
\Omega_m\rightarrow \infty$. Since $0\leq \theta\leq \frac{\pi}{2},$ then
$\alpha_1> 0$.
Since
$\left(\frac{d \theta'}{d
\theta},r'\right)|_{\theta=\frac{\pi}{2}}=\left(\frac{6}{
(1+\alpha_1^2)\alpha_2},-\frac{12}{(1+\alpha_1^2)\alpha_2}\right)$, we
deduce that $Q_2$ is
unstable for $\alpha_2>0$ or stable for $\alpha_2<0$, as confirmed in
Figure  \ref{fig2}. Since at this point $\Omega_m$
diverges, it corresponds to some form of future (respectively past)
singularity for $\alpha_2<0$ (respectively $\alpha_2>0$)
\cite{Sami:2003xv,Nojiri:2005sx,Briscese:2006xu,
Bamba:2008ut,Capozziello:2009hc,
Saridakis:2009jq}. Its detailed classification for the various parameter
regions lies beyond the scope of the present work.

\item
$Q_3: \theta^*=\frac{\pi}{2},  r^*=1, x=0, \Omega_m=\infty$. Since
$\left(\frac{d \theta'}{d
\theta},r'\right)|_{\theta=\frac{\pi}{2}}=\left(0,0\right)$, we cannot rely
on the linearization to examine the stability, and therefore, we need to resort
to numerical examination (see Figures \ref{fig1},
\ref{fig2}). Since at this point $\Omega_m$
diverges, it corresponds to some form of future, past or intermediate singularity
\cite{Sami:2003xv,Nojiri:2005sx,Briscese:2006xu,
Bamba:2008ut,Capozziello:2009hc,
Saridakis:2009jq}.

\end{itemize}
The above results are summarized in Table \ref{tab5}.
\begin{table*}[!]
\begin{center}
\begin{tabular}{|c|c|c|c|c|c|c|}
\hline
 Cr. P. & $\theta^*$ &
$\frac{d \theta'}{d\theta}\left|_{\theta=\theta^*}\right.$ & $r'
\left|_{\theta=\theta^*} \right.$ &
 Stability \\
\hline \hline
$Q_1$& $0$
& $-\frac{3}{2\alpha_2}$ &  $-\frac{3}{2\alpha_2}$ &   saddle point  \\
\hline
$Q_2$& $\arctan\left(\frac{\alpha_1}{2}\right)$
&   $\frac{6}{(1+\alpha_1^2)\alpha_2}$  &
$-\frac{12}{(1+\alpha_1^2)\alpha_2}$ &    unstable for $\alpha_2>0$ \\
& &  &       &stable for $\alpha_2<0$  \\
\hline
$Q_3$& $\frac{\pi}{2}$
&   $0$ &
$0$ &    see numerical elaboration \\
\hline
\end{tabular}
\end{center}
\caption[crit]{\label{tab5} The real critical points of the
autonomous system \eqref{eqx}-\eqref{eqm} at infinity, their existence
conditions, the corresponding values of $\frac{d \theta'}{d\theta}$ and
$r'$, and the resulting stability conditions.}
\end{table*}

\end{appendix}

\end{document}